# Fast particle-driven ion cyclotron emission (ICE) in tokamak plasmas and the case for an ICE diagnostic in ITER

K.G. McClements[1], R. D'Inca[2], R.O. Dendy[1,3], L. Carbajal[3], S.C. Chapman[3], J.W.S. Cook[3], R.W. Harvey[4], W.W. Heidbrink[5], S.D. Pinches[6]

[1] CCFE, Culham Science Centre, Abingdon, Oxfordshire OX14 3DB, UK
[2] Max-Planck-Institut für Plasmaphysik, Garching D-85748, Germany
[3] CFSA, Department of Physics, Warwick University, Coventry CV4 7AL, UK
[4] CompX, Del Mar, California 92014-5672, USA
[5] University of California, Irvine, California 92697, USA
[6] ITER Organization, Route de Vinon-sur-Verdon, CS 90 046, 13067 St. Paul-lez-Durance Cedex, France

E-mail: k.g.mcclements@ccfe.ac.uk

**Abstract**

Fast particle-driven waves in the ion cyclotron frequency range (ion cyclotron emission or ICE) have provided a valuable diagnostic of confined and escaping fast ions in many tokamaks. This is a passive, non-invasive diagnostic that would be compatible with the high radiation environment of deuterium-tritium plasmas in ITER, and could provide important information on fusion α-particles and beam ions in that device. In JET, ICE from confined fusion products scaled linearly with fusion reaction rate over six orders of magnitude and provided evidence that α-particle confinement was close to classical. In TFTR, ICE was observed from super-Alfvénic α-particles in the plasma edge. The intensity of beam-driven ICE in DIII-D is more strongly correlated with drops in neutron rate during fishbone excitation than signals from more direct beam ion loss diagnostics. In ASDEX Upgrade ICE is produced by both super-Alfvénic DD fusion products and sub-Alfvénic deuterium beam ions. The magnetoacoustic cyclotron instability (MCI), driven by the resonant interaction of population-inverted energetic ions with fast Alfvén waves, provides a credible explanation for ICE. One-dimensional particle-in-cell (PIC) and hybrid simulations have been used to explore the nonlinear stage of the MCI, thereby providing a more exact comparison with measured ICE spectra and opening the prospect of exploiting ICE more fully as a fast ion diagnostic. For realistic values of fast ion concentration, nonlinearly saturated ICE spectra in these simulations closely resemble measured spectra. The PIC/hybrid approach should soon make it possible to simulate the nonlinear physics of ICE in full toroidal geometry. Emission has been observed at a wide range of poloidal angles, so there is flexibility in the location of ICE detectors. Such a detector could be implemented in ITER by installing a small toroidally-orientated loop near the plasma edge or by adding a detection capability to the ICRH antennae.

## 1. Introduction

Electromagnetic emission from magnetically-confined plasmas in the ion cyclotron frequency range is generally categorised as ion cyclotron emission (ICE). This emission is invariably driven by ions that are superthermal but not necessarily super-Alfvénic; ICE from deuterium plasmas heated Ohmically or by hydrogen beams was reported in the early years of JET operation [1,2]; here the emission intensity scaled with the deuterium-deuterium neutron rate and for this reason was attributed to the charged products of thermonuclear fusion reactions. ICE is usually detected using either an ion cyclotron resonance heating (ICRH) antenna in receiver mode or a dedicated radio frequency probe (loop or dipole). Fusion experiments in which emission identified as ICE has been detected include ASDEX Upgrade [3], DIII-D [4], JET [1,2], JT-60U [5], LHD [6], PDX [7], and TFTR [8]. This is a passive, non-invasive diagnostic measurement that would be compatible with the high radiation environment of DT burning plasmas in ITER and could yield important information on the fusion α-particle and beam ion populations in that device, complementing other diagnostics such as collective Thomson scattering. In this paper we report on recent progress in the experimental (section 2) and theoretical (section 3) study of ICE, and present the case for an ICE diagnostic in ITER (section 4).

## 2. Observations of ICE

In the 1991 JET Preliminary Tritium Experiment (PTE), peaks in ICE intensity were observed in both pure deuterium and DT pulses at frequencies close to the α-particle/ deuterium cyclotron frequency $\omega_{cD}$ in the outer midplane edge (17MHz) and at cyclotron harmonics $n$ up to $n = 10$ [9]; higher harmonics appeared to merge into a continuum. A linear relation was found between ICE intensity and neutron rate in JET over six orders of magnitude. The total ICE power measured using the JET ICRH antenna (a few μW at most) was many orders of magnitude smaller than the radio frequency power $P_{RF}$ (several MW) which the antenna was capable of launching into the plasma. ICE detected in the 1997 DT experiments in JET provided evidence of classical α-particle confinement [10].

ICE was also detected in JET using a dedicated probe mounted on the inner side of vacuum vessel, in particular during (H)D minority ICRH [11]. In these pulses the minority protons typically had a temperature perpendicular to the magnetic field of around 2 MeV. The lower plot in figure 1 shows the radio frequency (RF) spectrum during a pulse with $P_{RF} = 5.6$ MW, in which four radio frequency generators were tuned to frequencies clustered around 50 MHz. In addition to peaks associated with ICRH, two additional peaks corresponding to the hydrogen cyclotron frequency $\omega_{cH}$ and its second harmonic in the outer midplane (see upper plot) are clearly visible. A key point here is that the very large disparity between the transmitted and received RF power does not prevent ICE from being detected. More recently, ICE driven by $^3$He minority fast ions (accelerated by ICRH) was detected using a sub-harmonic arc detection system on the JET ICRH antennas [12].

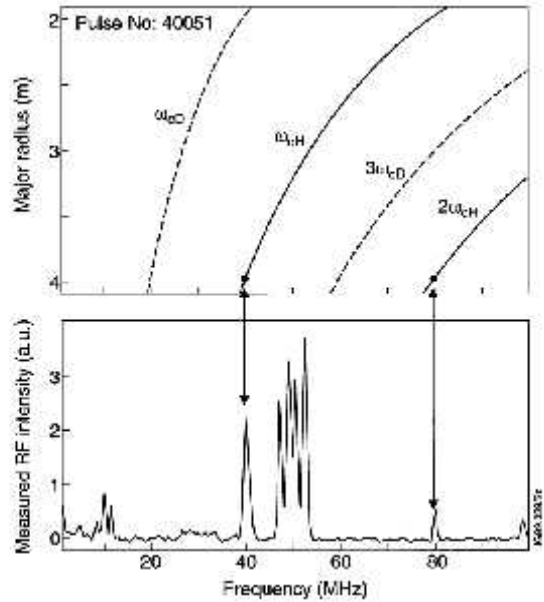

**Figure 1.** Lower plot: radio frequency emission spectrum during four-frequency H minority ICRH in JET pulse 40051. ICE peaks corresponding to the H cyclotron frequency and its second harmonic in the outer plasma edge (indicated by the upper plot) can be seen [11].

In recent years ICE has been observed in ASDEX Upgrade [3], DIII-D [4], JT-60U [5] and LHD [6]. In ASDEX Upgrade ICE is detected using a dedicated probe (consisting of two cross-dipole antennae) and a voltage probe inside the ICRH antenna on the high and low field sides of the plasma respectively. ICE has been observed at $\omega_{cD}$ in the plasma centre (plus its second and third harmonics) during deuterium NBI and at $\omega_{cH}$ in the outer plasma edge during minority hydrogen ICRH. In pulses with high power (> 5MW) deuterium beam injection into deuterium target plasmas, transient ICE has been detected at the second harmonic of the $^3$He cyclotron frequency in the plasma edge (probably driven by $^3$He fusion products), along with longer-lasting emission at several deuterium harmonics (figure 2). While $^3$He fusion products are born with speeds $v$ significantly greater than the Alfvén speed $c_A$ in these plasmas, deuterium beam ions have $v$ of the order of $c_A$ or less, from which one may conclude that ICE does not require the fast ion population to be strongly super-Alfvénic. This was also found to be the case in TFTR [13].

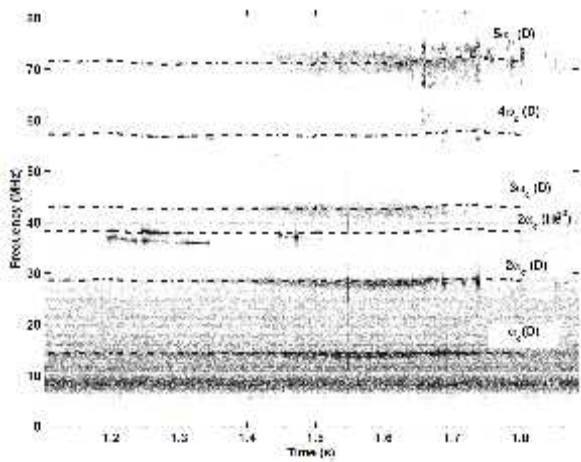

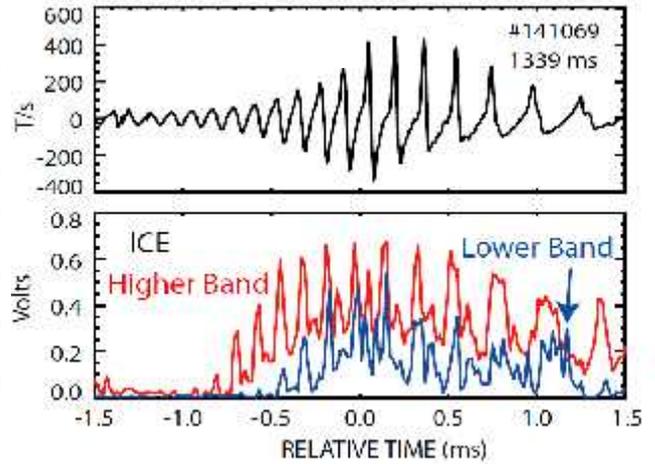

**Figure 2.** High field side spectrogram from ASDEX Upgrade discharge 26915, a D plasma with toroidal field 2.5T, central density $6\times10^{19}$m$^{-3}$ and NBI power 10MW. The dashed lines indicate cyclotron harmonics of D/H and $^3$He at normalised minor radius $r/a = 0.95a$.

**Figure 3.** Mirnov coil signal (top) and ICE signals in two frequency bands (bottom) during fishbone in DIII-D. The lower and higher bands include, respectively, the deuterium cyclotron frequency at the plasma edge and the second and third harmonics of this [4].

In DIII-D a magnetic loop on the low field side [14] filters the ICE signal into a lower band, which includes $\omega_{cD}$ at the plasma edge, and a higher band which covers $2\omega_{cD}$ and $3\omega_{cD}$. Bursting beam ion-driven ICE is observed in both frequency bands during fishbone excitation, but with a phase lag relative to the fishbone signal (figure 3) [4]. In this case the ICE appears to be driven by the fishbone-induced expulsion of beam ions to the plasma edge; as in other conventional tokamaks, the beam ions are sub-Alfvénic. In fact ICE in DIII-D is found to be correlated more strongly with drops in neutron flux during fishbone excitation than are signals from more direct measurements of beam ion losses.

## 3. Interpretation of ICE

ICE is generally attributed to the magnetoacoustic cyclotron instability (MCI), which results from the resonant interaction of energetic ions with fast Alfvén waves [15]. Instability can be driven by anisotropy, population inversion perpendicular to the magnetic field or radial gradients. In the case of JET, inverted fusion α-particle distributions of the type required to drive ICE arose in the outer midplane plasma edge due to the large radial excursions from the core to the edge of marginally-trapped particles [9]. The local α-particle distribution in the outer midplane $f_\alpha$ was dominated by particles originating from the plasma core, and consequently had a bump-in-tail determined by the fusion reaction rate in the core and the usual constants of motion. Waves propagating obliquely with respect to the magnetic field **B** have been shown to be linearly unstable when the energetic ion concentration is very low, due to the Doppler shift term in the cyclotron resonance condition causing thermal cyclotron damping to be decoupled from the drive. Nevertheless the growth rates in a uniform equilibrium plasma are typically found to peak at propagation angles that are nearly perpendicular to **B** [15]. The inclusion of grad-*B* and curvature drift terms in the cyclotron resonance condition leads to prediction of higher growth rates, and the maximum drive is again found for nearly perpendicular propagation [16]. Toroidal effects on the linear stability of these modes are discussed further in [17]. Generally, it is found that the linear drive falls off fairly rapidly (but remains finite) as the energetic ion distribution broadens or the Alfvénic Mach number falls below one.

It should be noted that there is no essential physical difference between ICE eigenmodes and compressional Alfvén eigenmodes (CAEs) in the ion cyclotron frequency range, which have been observed in both spherical [18,19] and conventional [20] tokamaks and, like ICE, have been attributed to fast Alfvén waves driven unstable due to cyclotron resonances with energetic ions. The close connection between ICE and CAEs is particularly illustrated by recent experiments at low toroidal magnetic field in the MAST spherical tokamak, where multiple CAEs were excited by deuterium beam ions in the ion cyclotron range [19]. In the spectrum shown in figure 4, three peaks occurring at frequencies close to $\omega_{cD}$ (evaluated at the magnetic axis) have amplitudes exceeding those of neighbouring peaks by a factor of $10^2$ or more.

The identification of ICE (and CAEs) with fast Alfvén waves has prompted several investigations of the global structure of these modes in tokamak geometry. Gorelenkov and Cheng [21] developed an analytical theory of fast Alfvén eigenmode structure for circular cross-section, finite aspect ratio cold plasmas. The modes were found to be localised radially to minor radii $r = r_0$ such that

$$\frac{r_0^2}{a^2} = \frac{1}{1+\sigma} - (2s+1)\frac{\sqrt{2\sigma/(1+\sigma)}}{m(1+\sigma)}, \qquad (1)$$

where *m*, *s* are poloidal and radial mode numbers, and σ is a parameter controlling the density profile *n(r)*:

$$n(r) = n_0\left(1 - r^2/a^2\right)^\sigma . \qquad (2)$$

For flat profiles (σ < 1), ICE eigenmodes are thus predicted to be localised near the plasma edge. However eigenfunctions of modes in the ion cyclotron range computed numerically by Smith and Verwichte [22] for a range of plasma shapes (but still using a cold plasma model) have a more extended radial structure.

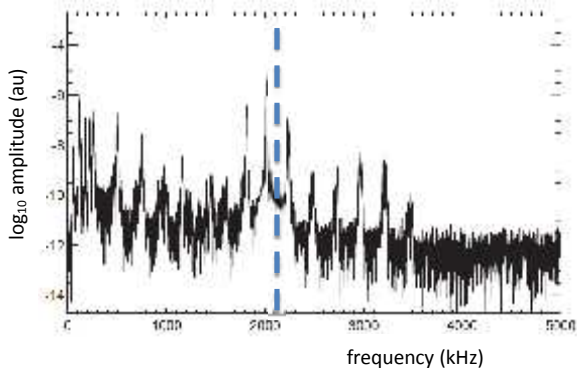

**Figure 4.** Spectrum of fluctuations measured using Mirnov coil in deuterium beam-heated MAST pulse 27147. The dashed line indicates $\omega_{cD}/2\pi$ at the magnetic axis [19].

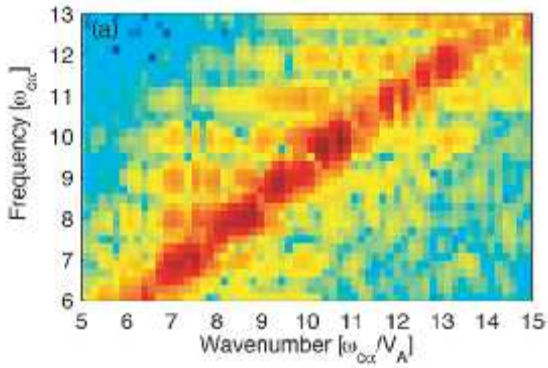

**Figure 5.** Wavenumber-frequency plot for the oscillatory component of B in a PIC simulation of the MCI, showing excitation of multiple $\alpha$-particle cyclotron harmonics [23].

There has been significant recent progress in the nonlinear modelling of ICE [23,24]. Particle-in-cell (PIC) simulations with one space and three velocity dimensions have been used, for the first time, to study self-consistently the excitation of the MCI by a population-inverted fast ion distribution and the subsequent relaxation of this distribution [23]. Specifically, the initial fast ion distribution ($\alpha$-particles) was modelled as a monoenergetic gyrotropic ring. Fourier analysis of the results in the space and time domains clearly shows the excitation of waves on the fast Alfvén branch, with the strongest growth occurring close to $\alpha$-particle cyclotron harmonics (figure 5). Hybrid simulations, again in one space dimension, which track full orbits of ions and treat electrons as a neutralising fluid, have been used to explore the later stages of the MCI [24], thereby providing a more exact comparison with measured ICE spectra and opening the prospect of exploiting more fully this emission process as a fast ion diagnostic. The initial fast ion distribution was again modelled as a monoenergetic gyrotropic ring. In simulations with parameters close to those of the PTE in JET, the nonlinearly-saturated ICE amplitude was found to depend strongly on the fast ion concentration $\xi$ at low cyclotron harmonics, but only weakly on $\xi$ at high harmonics (figure 6). For realistic $\xi$ the nonlinearly-saturated spectrum closely resembles the measured one. Using this hybrid approach in combination with a finite orbit width Fokker-Planck code such as CQL3D to generate realistic fast ion distributions [25], it will soon become possible to simulate nonlinearly ICE in full toroidal geometry.

## 4. Excitation and detection of ICE in ITER plasmas

As an illustrative example, we consider an ITER plasma in the baseline inductive mode of operation with toroidal field $B_0 = 5.3$T, electron density $n_e = 10^{20}$m$^{-3}$, equal concentrations of deuterium and tritium, 2% helium, 2% beryllium and 0.1% argon [26]. In this case the Alfvén speed $c_A$ will be approximately (5-6)$\times 10^6$ms$^{-1}$ on the low field side of the plasma. As noted previously, ICE can be strongly driven by ions with speeds $v$ in excess of $c_A$. Depending on velocity-space gradients, fusion $\alpha$-particles and 1MeV beam deuterons, born respectively with $v = 1.3\times10^7$ ms$^{-1}$ and $v = 10^7$ ms$^{-1}$, are thus expected to provide strong drive for ICE in ITER, particularly since they will have much higher densities than α-particles in JET DT plasmas. However, compared to JET or TFTR, fusion α-particles in ITER will have smaller orbits relative to the machine size and, in the absence of significant levels of non-classical transport, will be better confined; beam ions are also expected to be well-confined. Centrally-born trapped fusion α-particles will have a potato orbit width $\Delta_r$ given by [10]

$$\frac{\Delta_r}{a} \approx 8.8\left(\frac{a}{R}\right)^{1/3}\left(\frac{m_r v}{Z_r e \mu_0 I_p}\right)^{2/3}, \quad (3)$$

where $R$ and $a$ are major and minor radii, $m_r$ and $Z_r e$ are the α-particle mass and charge, $I_p$ is the plasma current and $\mu_0$ is free space permeability. This expression yields $\Delta_r = 0.4a$ and $0.5a$ for $I_p = 15$MA and 10MA respectively. Trapped α-particles born further out will have a maximum orbit width [27]

$$\frac{\Delta_r}{a} \approx \frac{a}{(Rr)^{1/2}}\frac{2fm_r v}{Z_r e \mu_0 I_p}. \quad (4)$$

When $r = a/2$ this expression gives $\Delta_r = 0.07a$ and $0.1a$ for $I_p = 15$MA and 10MA. Given that the α-particle density profile $n_r$ in both scenarios is predicted to be strongly peaked in the plasma centre, it may be inferred from these estimates that ring-like energetic ion distributions of the type believed to have driven ICE in JET are unlikely to be found in the ITER edge plasma under steady state conditions, although they could arise deeper inside the plasma, depending on how steeply $n_r$ falls off with radius.

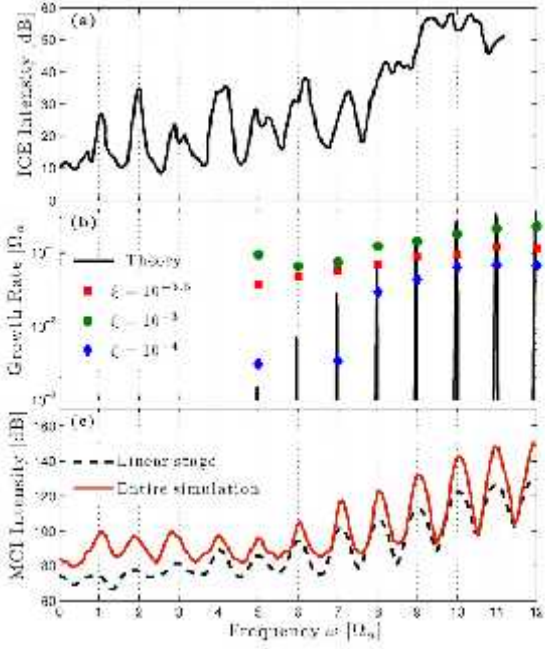

**Figure 6.** (a) Measured ICE spectrum in JET DT pulse 26148. (b) MCI linear growth rates from analytical theory (for $\xi = 10^{-3}$: black curve) and hybrid simulations (for three values of $\xi$: coloured points). (c) $B_z$ intensity in linear (dashed curve) and nonlinear (solid curve) stages of instability in hybrid simulation with $\xi = 10^{-3}$ [22].

The steady-state excitation of ICE in ITER would depend also on the structure of the corresponding Alfvén eigenmodes, and, if excited deep inside the plasma, it is unclear that ICE could propagate to a detector without being strongly absorbed *en route*, for example due to the ion-ion hybrid resonances that exist in DT plasmas [28]. However, even if detectable levels of ICE are not normally produced in steady state conditions in ITER, it is likely that it could be used as an alternative method of studying fast ion redistribution and losses due to MHD activity. In addition to the correlations between ICE and fishbones observed in DIII-D [4], clear evidence has also been found of links between this type of emission and both sawteeth and ELMs in JET [9] and with TAEs in LHD [6]. The detection of ICE in ITER would provide information on energetic ion behaviour supplementing that obtained using other diagnostics, such as collective Thomson scattering and γ-ray detectors. It should be noted that the heat loads in the DT phase of ITER operation will make it impossible to use a conventional fast ion loss detector unless a reciprocating drive system is used [29]. ICE detection could provide an alternative method of studying -particle and beam ion losses.

In principle detection of ICE is possible using any technique for measuring magnetic, electric or density fluctuations in the ion cyclotron range. Dedicated radio frequency probes have been used at a wide range of poloidal locations, for example at the top of the vessel in TFTR [30] and the high field side in JET [11]. The design requirements of an ICE diagnostic system for ITER are thus flexible. Other detection possibilities include microwave reflectometry, which is planned for ITER [31] and has been used to measure the spatial structure of modes at frequencies up to the CAE range in NSTX [32]; this is not a passive technique, however, since it requires microwaves to be launched into the plasma. Generally, ICE detection in existing devices is a passive, non-invasive diagnostic technique. It should be fully compatible with the high radiation environment of DT operation in ITER. For a dedicated diagnostic, the detector could be a modified plasma-facing component, as in DIII-D [14]; this works well for modes with predominately compressional polarization, such as ICE. If possible, the detector should be capable of detecting frequencies ranging up to around the tenth harmonic of the α-particle cyclotron frequency, i.e. several hundred MHz. It would be particularly useful to have a toroidally-distributed array of detectors, since this would provide wavevector information as well as spectra and thereby provide additional information on the fast ion distribution that is exciting the emission. ICE measurements in JET and ASDEX Upgrade show that a dedicated ICE probe can operate well during ICRH [3,11], in particular that the ICE signal can be clearly distinguished from that due to the ICRH source, despite the very large disparity noted previously between ICRH power coupled to the plasma and the measured power in ICE.

Alternatively, or in addition to a dedicated diagnostic, a detection capability could be added to the ITER ICRH antenna system (for a technical description of this system, see [33]). ICE measurements could be carried out with the antenna in passive reception mode, as in the JET PTE experiment [9]. However the recent measurements in ASDEX Upgrade [3] demonstrate that it is also possible to detect emission with ICRH antennae during active operation of the same antennae for plasma heating: this in fact was how ICE in ASDEX Upgrade was first observed, with a probe located inside the ICRH transmission line. For this system to work it was necessary to ensure that the probe was located near a maximum voltage point of the standing wave pattern in the transmission line. The frequency of the ICRH source needed to be eliminated from the signal using a filter, with strong attenuation in a narrow bandwidth around the rejected frequency. A low-noise amplifier was also required to extract the ICE signal from the background. In a future system of this type on ITER, wavevector information could be determined by comparing the phase of the emission in different straps of the ICRH antennae; as noted above, such measurements are likely to enhance significantly the utility of ICE as a fast particle diagnostic. On ASDEX Upgrade it was found that measurements obtained using the dedicated ICE probe had a somewhat better signal-to-noise ratio than those obtained using the probe in the ICRH system.

## 5. Summary

ICE was used to obtain valuable information on the behaviour of fusion α-particles in the two large tokamaks capable so far of DT operation (JET and TFTR), and continues to be a useful diagnostic of confined and escaping fast ions in many fusion experiments. This is a passive, non-invasive diagnostic that would be compatible with the high radiation environment of DT plasmas in ITER, and could thus provide a valuable additional route to the experimental study of fusion α-particles and beam ions in that device. In JET, ICE from confined fusion products scaled linearly with the fusion reaction rate over six orders of magnitude, and provided evidence that α-particle confinement was close to classical. More recently, the intensity of beam-driven ICE in DIII-D has been found to be more strongly correlated with drops in neutron rate during fishbone excitation than are signals from more direct beam ion loss diagnostics. ICE in ASDEX Upgrade has been observed to be excited by both super-Alfvénic DD fusion products and sub-Alfvénic deuterium beam ions. The MCI, driven by the resonant interaction of population-inverted energetic ions with fast Alfvén waves, provides a credible explanation for ICE. PIC and hybrid simulations have been used to explore the nonlinear stage of the MCI, thereby providing a more exact comparison with measured ICE spectra and opening the prospect of exploiting ICE more fully as a fast ion diagnostic; it should soon become possible to simulate nonlinear physics of ICE in full toroidal geometry. Emission has been observed at a wide range of poloidal locations, so there is considerable flexibility in the requirements of an ICE detector. Such a system could be implemented in ITER by installing a dedicated probe, for example a magnetic loop, or by adding a detection capability to the ICRH antenna system. In the latter case measurements on ASDEX Upgrade demonstrate the feasibility of using antennae simultaneously to heat plasma and to detect ICE.

**Acknowledgments**

This project has received funding from the European Union's Horizon 2020 research and innovation programme under grant agreement number 633053 and from the RCUK Energy Programme [grant number EP/I501045]. To obtain further information on the data and models underlying this paper please contact PublicationsManager@ccfe.ac.uk. The views and opinions expressed herein do not necessarily reflect those of the European Commission.
   The views and opinions expressed herein do not necessarily reflect those of the ITER Organization.